\documentclass[]{aa}

\def\rmit#1{{\it #1}}              
\def\specchar#1{{\sc #1}}

\def\CaIIH{\mbox{Ca\,\specchar{ii}\,\,H}}       
\def\CaII{\mbox{Ca\,\specchar{ii}}}

\def\eg{\rmit{e.g.}}

\usepackage{booktabs}
\usepackage{graphicx}
\usepackage{multirow}
\usepackage{color}
\usepackage[flushleft]{threeparttable}

\usepackage{array}
\newcolumntype{?}{@{\vrule width 2pt}}

\usepackage{hyperref}
\hypersetup{
    final=true,
    pageanchor=true,
    colorlinks=true,
    breaklinks=true,
    linkcolor=blue,
    citecolor=blue,
    urlcolor=blue,
    pdfpagemode=UseNone,
    pdftitle={Inversions of synthetic umbral flashes: effects of the scanning time on the inferred atmospheres},
    pdfauthor={T. Felipe et al.},
    pdfsubject={Solar Physics},
    pdfkeywords={Methods: numerical -- Sun: chromosphere -- sunspots -- Sun: oscillations -- Techniques: polarimetric}}

\titlerunning{Effects of the scanning time on inferred umbral flash atmospheres}   

\begin{document}



\title{Inversions of synthetic umbral flashes: effects of the scanning time on the inferred atmospheres}

\author{T. Felipe\inst{\ref{inst1},\ref{inst2}}
\and H. Socas-Navarro\inst{\ref{inst1},\ref{inst2}}
\and D. Przybylski\inst{\ref{inst3}}
}


\institute{Instituto de Astrof\'{\i}sica de Canarias, 38205, C/ V\'{\i}a L{\'a}ctea, s/n, La Laguna, Tenerife, Spain\label{inst1}
\and 
Departamento de Astrof\'{\i}sica, Universidad de La Laguna, 38205, La Laguna, Tenerife, Spain\label{inst2} 
\and 
Max-Planck-Institut f\"{u}r Sonnensystemforschung, Justus-von-Liebig-Weg 3, 37077 G\"{o}ttingen, Germany\label{inst3} 
}

\abstract
{The use of instruments that record narrow band images at selected wavelengths is a common approach in solar observations. They allow the scanning of a spectral line by sampling the Stokes profiles with two-dimensional images at each line position, but require a compromise between spectral resolution and temporal cadence. The interpretation and inversion of spectropolarimetric data generally neglect the changes in the solar atmosphere during the scanning of the line profiles.} 
{We evaluate the impact of the time-dependent acquisition of different wavelengths on the inversion of spectropolarimetric profiles from chromospheric lines during umbral flashes.}
{Numerical simulations of non-linear wave propagation in a sunspot model were performed with the code MANCHA. Synthetic Stokes parameters in the \CaII\ 8542 \AA\ line in NLTE were computed for an umbral flash event using the code NICOLE. Artificial profiles with the same wavelength coverage and temporal cadence from reported observations were constructed and inverted. The inferred atmospheric stratifications were compared with the original simulated models.}
{The inferred atmospheres provide a reasonable characterization of the thermodynamic properties of the atmosphere during most of the phases of the umbral flash. Only at the early stages of the flash, when the shock wave reaches the formation height of the \CaII\ 8542 \AA\ line, the Stokes profiles present apparent wavelength shifts and other spurious deformations. These features are misinterpreted by the inversion code, which can return unrealistic atmospheric models from a good fit of the Stokes profiles. The misguided results include flashed atmospheres with strong downflows, even though the simulation exhibits upflows during the umbral flash, and large variations in the magnetic field strength.}
{Our analyses validate the inversion of Stokes profiles acquired by sequentially scanning certain selected wavelengths of a line profile, even in the case of rapidly-changing chromospheric events such as umbral flashes. However, the inversion results are unreliable during a short period at the development phase of the flash.}

\keywords{Methods: numerical -- Sun: chromosphere -- sunspots -- Sun: oscillations -- Techniques: polarimetric}

\maketitle


\section{Introduction}

Spectropolarimetry is the most powerful method for inferring the thermodynamic and magnetic properties of the solar atmosphere. Two different approaches are used for data acquisition: two-dimensional (2D) instruments, such as Fabry-P\'erot and Michelson interferometers, and slit-spectropolarimeters. Each of these approaches has advantages and drawbacks. Two-dimensional instruments allow the measurement of a 2D field of view in narrow-band images across the profile of the spectral line of interest \citep[\eg,][]{Tritschler+etal2002, Cavallini2006, Puschmann+etal2006, Scharmer+etal2008, Schou+etal2012}. The 2D images at each wavelength are obtained successively at different times and, thus, the observer needs to look for a compromise between the spectral sampling (the number of wavelength positions scanned) and the temporal cadence of the observations. On the contrary, slit-spectropolarimeters \citep[\eg,][]{Tsuneta+etal2008, Collados+etal2012, dePontieu+etal2014} provide spectra with a much better spectral resolution and with all the wavelengths obtained simultaneously. However, only a one dimensional slice of the field of view is acquired at any given time. This means that observing a 2D region of the solar surface requires the construction of a raster map by shifting the slit in small increments perpendicular to its direction. In this case, the retrieved 2D map does not correspond to a single moment, but its measurement spans during a certain temporal range.

In the case of 2D instruments, the measured line profiles are composed by the signals acquired sequentially in different wavelengths. Since the solar atmosphere is a time-varying medium, the retrieved profiles do not characterize the atmosphere at a certain time, but they represent a mixture of the spectra generated over a temporal period instead. However, the interpretation and inversion of spectropolarimetric data generally neglect this effect. For slow evolving processes, whose characteristic time is much longer than the scanning time, this is a reasonable assumption. For highly dynamic phenomena, such as flares \citep{Guglielmino+etal2016,Kleint2017} or umbral flashes \citep[UFs,][]{delaCruz-Rodriguez+etal2013, Henriques+etal2017}, it can potentially lead to a misinterpretation of the polarimetric profiles.  

The aim of this work is to evaluate the signatures of rapid changes of the solar atmosphere on the spectropolarimetric line profiles acquired by 2D instruments. We have chosen to focus on UFs. They were first detected by \citet{Beckers+Tallant1969} as sudden brightenings in \CaII\ H and K lines in the umbra of sunspots. These authors proposed that umbral flashes were produced by magneto-acoustic wave propagation, and this suggestion was confirmed by \citet{Havnes1970}, who found that the flash in \CaII\ lines takes place during the compressional stage of a magneto-acoustic wave, which produces an increase in the temperature and the number density of \CaII\ atoms. This way, UFs are one of the many manifestations of wave propagation in sunspots \citep[see][for a review]{Khomenko+Collados2015}. The umbral chromosphere is dominated by oscillations with frequency around 5-6 mHz. These waves come directly from the photosphere propagating along field lines. Their amplitude increases with height faster than that of evanescent waves (frequency below $\sim 4$ mHz) and steepen into shocks at the chromosphere \citep{Centeno+etal2006, Felipe+etal2010b}. Slow mode waves propagating from the photosphere to higher atmospheric layers along the inclined field lines of the penumbra generate a visual pattern of radially outward chromospheric waves known as running penumbral waves \citep{Bloomfield+etal2007b, Madsen+etal2015}. Recently, the detection of the photospheric counterpart of the running penumbral waves has been reported \citep{LohnerBottcher+BelloGonzalez2015, Zhao+etal2015}.

\citet{SocasNavarro+etal2000a,SocasNavarro+etal2000b} performed the first polarimetric observations of UFs. They found anomalous Stokes $V$ profiles, which were interpreted as the contribution of two unresolved atmospheres. Each pixel is partially occupied with a component at rest or with a small downflow and the rest is filled with a shock wave, including a strong upflow velocity and producing line core emission reversal. The unresolved structure was inferred by another independent observation \citep{Centeno+etal2005} and then directly observed in Hinode imaging observations \citep{SocasNavarro+etal2009}. Recent observations \citep[\eg,][]{delaCruz-Rodriguez+etal2013} are starting to resolve the two components. \citet{delaCruz-Rodriguez+etal2013} reported NLTE inversions of high-resolution spectropolarimetric \CaII\ 8542 \AA\ observations of UFs. They inferred umbral temperature enhancements of up to 1000 K in the flashed atmosphere and penumbral magnetic field oscillations with peak-to-peak amplitude around 200 G associated to the running penumbral waves. Recently, \citet{Henriques+etal2017} inverted a large set of UFs and found two different families of solutions: some observed flashed profiles were characterized by models with upflows, in agreement with \citet{SocasNavarro+etal2000a,SocasNavarro+etal2000b} and \citet{delaCruz-Rodriguez+etal2013}; while others were better reproduced by atmospheres with strong downflows, close to the sound speed.

Using NLTE hydrodynamic simulations, \citet{Bard+Carlsson2010} were able to reproduce the intensity profiles in \CaIIH\ of the flashed atmosphere as the result of acoustic waves generated by a photospheric piston which steep into shock in the chromosphere. \citet{Felipe+etal2014b} performed the synthesis of the spectropolarimetric profiles of the \CaII\ infrared triplet produced by UFs in observationally driven numerical simulations \citep{Felipe+etal2011}. In this work, we have followed the approach from \citet{Felipe+etal2014b}, but the spectropolarimetric synthesis has been performed taking into account the time difference in the data acquisition by 2D instruments. The paper is organized as follows: the numerical methods, including the magnetohydrodynamic (MHD) code and the synthesis and inversion code, are described in Sect. \ref{sect:methods}, the results are presented in Sect. \ref{sect:results}, and the conclusions are summarized in Sect. \ref{sect:conclusions}.

\begin{figure}[!ht] 
 \centering
 \includegraphics[width=9cm]{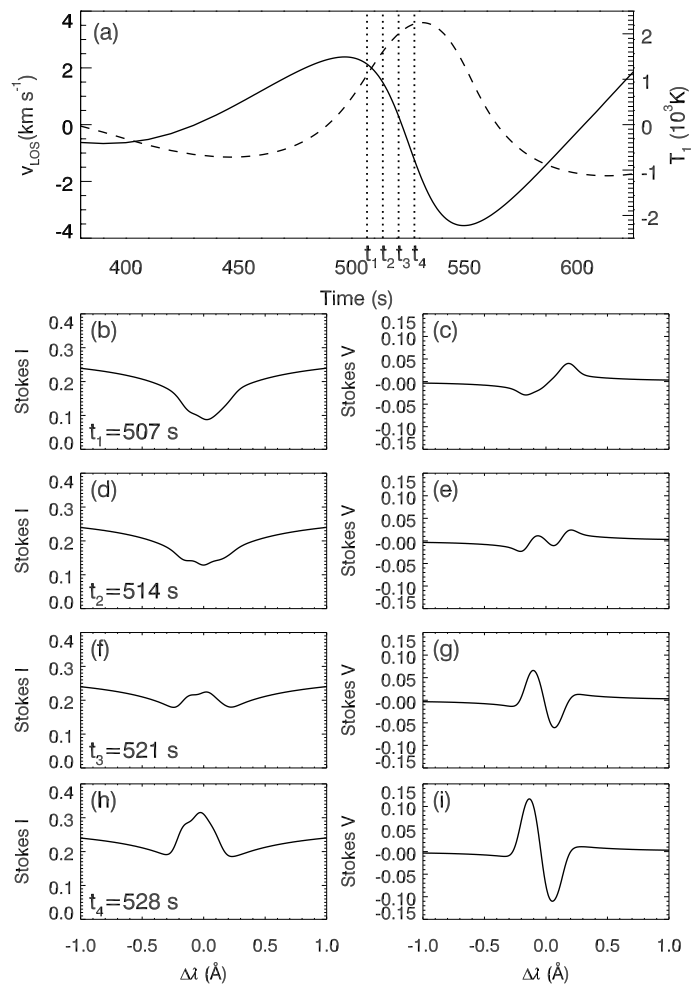}
  \caption{Spectropolarimetric synthesis of the \CaII\ 8542 \AA\ line at several time steps. Top panel: temporal evolution of the vertical velocity (solid line, left axis) and temperature perturbation (dashed line, right axis) at constant geometrical height \hbox{$z=750$ km}. Rest of the panels: Stokes I (left panels) and Stokes V (right panels) synthesized at a fixed time step. The time is indicated at the top left side of the Stokes I figures and by the vertical dotted line in the top panel.}
    
  \label{fig:sintesis_tfijo}
\end{figure}

\section{Numerical methods}
\label{sect:methods}

We have simulated non-linear wave propagation from the photosphere to higher atmospheric layers in a sunspot using the code MANCHA \citep{Khomenko+Collados2006, Felipe+etal2010a}. Waves are driven below the solar surface as perturbations to a magnetohydrostatic background model. They develop into shocks when reaching the chromosphere. The simulation provides a reasonable characterization of the solar atmosphere during UF events. Its details can be found in the Appendix \ref{appendix:simulations}. In the following, we will focus on the analysis of a single, time dependent atmosphere, which corresponds to a single column of the MHD simulation.

\begin{figure}[!ht] 
 \centering
 \includegraphics[width=9cm]{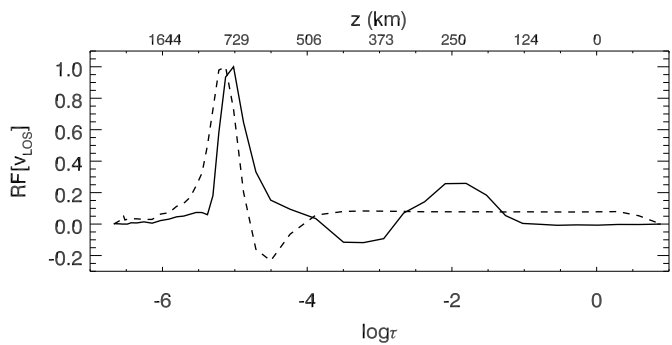}
  \caption{Response functions of the intensity of the \CaII\ 8542 \AA\ line to the LOS velocity at 8541.94 \AA\ for the umbral atmosphere at rest (solid line) and VAL-C model (dashed line).}    
  \label{fig:RF}
\end{figure}

The four Stokes profiles for the \CaII\ 8542 line produced by the model at an umbral position (at 0.8 Mm from the sunspot center, where a UF is produced after approximately 5 min of simulation) have been synthesized (and inverted) using the NLTE code NICOLE \citep{SocasNavarro+etal2015}. NICOLE assumes complete angle and frequency redistribution and a plane-parallel atmosphere for each pixel. In the presence of magnetic fields, the code computes the polarization induced by the Zeeman splitting. The \CaII\ atom model used includes five bound levels and the continuum \citep{delaCruz-Rodriguez+etal2012}. The line broadening due to collisions was estimated following the formalism of \citet{Anstee+O'Mara1995}. The NLTE capabilities of NICOLE have been validated using MHD numerical simulations up to the chromosphere \citep{delaCruz-Rodriguez+etal2012}. The core of the synthesized \CaII\ 8542 line is narrower than that measured from observational data, due to the lack of small scale random motions in the simulation. More realistic synthetic profiles have been obtained by including an artificial microturbulent velocity in the synthesis, similar to \citet{delaCruz-Rodriguez+etal2012}. A microturbulence of 3 km s$^{-1}$ was found sufficient for producing \CaII\ 8542 intensity profiles with a FWHM comparable to those reported from observations \citep[\eg][]{delaCruz-Rodriguez+etal2013,Henriques+etal2017}.

In the inversion mode, the stratification of the atmospheric physical variables is inferred through an iterative process in which an initial guess atmosphere is modified until a good match (defined as an absolute minimum in the merit function $\chi^2$) between the observed (synthesized in our case) profiles and those produced by the resultant atmosphere is found. We have used the FAL-C model \citep{Fontenla+etal1993} as initial atmosphere, and two cycles were employed for the computation of the inversion. The first cycle generates an approximate solution from the use of a small amount of nodes per atmospheric quantity. This solution is then refined through a new inversion with a higher number of nodes during the second cycle. In both cases, the nodes are uniformly spaced in $\log\tau$. The distribution of nodes is detailed in Table \ref{table:nodes}. Different weights were assigned to each of the Stokes parameters. The weight of Stokes $V$ was 70$\%$ of that from Stokes $I$, whereas Stokes $Q$ and $U$ have a weight of 50$\%$. Although Stokes $Q$ and $U$ signals are very weak, we have included them in the inversion in order to constrain the solutions generated by NICOLE to profiles with low amplitude in linear polarization. All the inversions were repeated 50 times with randomized initializations and the solution with lowest $\chi^2$ was selected. This method reduces the chances that the algorithm settles into a local minimum in $\chi^2$. In order to estimate the uncertainties of the method, this process (selecting the best inversion from 50 cases with different initializations) has been repeated five times. The atmospheric stratification is taken as the case with lowest $\chi^2$ (over 250 cases), while the error is estimated as the standard deviation over the five independent realizations of 50 inversions.

\begin{table}
\begin{center}
\caption[]{\label{table:nodes}
          {Summary of the nodes used for the inversion.}}
\begin{tabular}{ccc}
\hline\noalign{\smallskip}
Physical variable 	& 	Nodes first cycle	& Nodes second cycle	\\
\hline\noalign{\smallskip}
Temperature		& 	4			& 8			\\
LOS velocity		& 	2			& 5			\\
$B_{\rm z}$		& 	1			& 2			\\
$B_{\rm x}$		& 	0			& 1			\\
$B_{\rm y}$		& 	0			& 1			\\
Microturbulence		& 	0			& 1			\\
\hline

\end{tabular}
 
\end{center}
\end{table}

\section{Results}
\label{sect:results}

In this work, we focus on the analysis of a single UF event from the simulations. As a first step, the synthesis has been performed for all the wavelengths at the same time, producing the ``instantaneous profiles'' (Sect. \ref{sect:fix_time}). Then, we show the synthesized Stokes profiles taking into account temporal limitations of 2D instruments with different configurations (Sect. \ref{sect:tstep}). We will refer to those as ``scanned profiles''.

\subsection{Synthetic instantaneous observations}
\label{sect:fix_time}

\subsubsection{Stokes profiles}

Figure \ref{fig:sintesis_tfijo} illustrates the Stokes I and V profiles of the \CaII\ 8542 \AA\ line at four time steps during the development of an UF. The top panel shows the temporal evolution of the vertical velocity and the temperature perturbation ($T_{\rm 1}(z,t)=T(z,t)-T_{\rm 0}(z)$, where $T_{\rm 0}$ is the temperature stratification of the background atmosphere) at a fixed geometrical height corresponding to the maximum of the response function of the \CaII\ line to the line-of-sight (LOS) velocity (see Fig. \ref{fig:RF}). The vertical velocity shows a peak-to-peak amplitude around \hbox{6 km s$^{-1}$} with the characteristic saw-tooth profile of shock propagation \citep[\eg,][]{Centeno+etal2006}. At the same geometrical height, the magnetic field shows small amplitude oscillations, with peak-to-peak amplitudes below 2.8 G.

The chosen time steps for the bottom panels sample the sudden change in the vertical velocity produced by the pass of the shock wave. The Stokes profiles undergo profound changes, including the appearance of the blueshifted emission core that characterize UFs (\hbox{$t=521$ s}, panel f), and a polarity reversal in Stokes $V$ (panel g). These changes take place during a short time period. The evolution from a quiescent profile to a flashed profile occurs in less than \hbox{14 s}. This temporal window is below the usual time employed by 2D instruments to complete the spectropolarimetric scan of a single spectral line.

\begin{figure*}[!ht] 
 \centering
 \includegraphics[width=18cm]{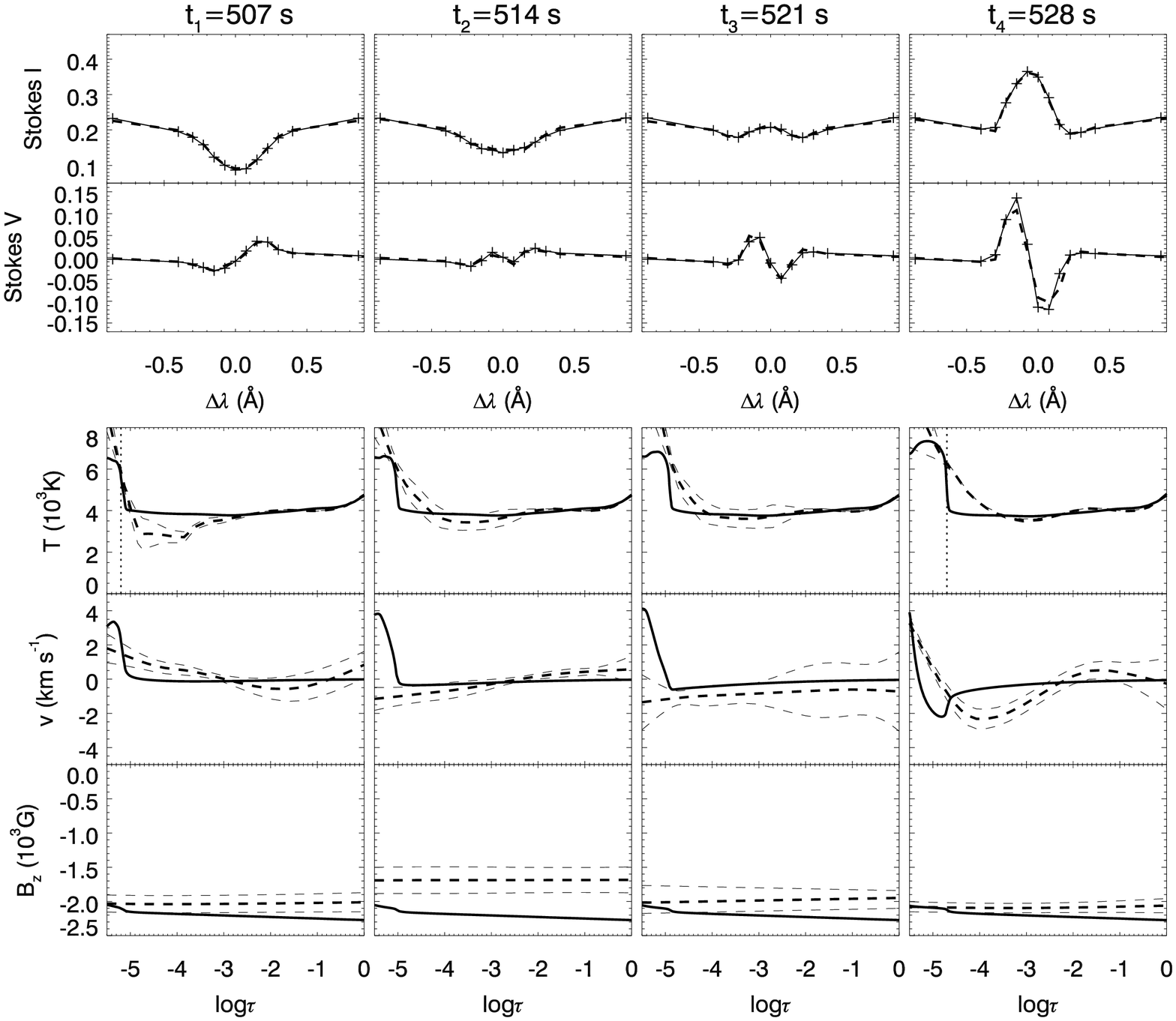}
  \caption{Top two rows: Synthesized instantaneous (solid lines with crosses at the line positions considered) and inverted (dashed lines) Stokes $I$ and $V$ profiles at four time steps (columns) during the development of an UF. Bottom three rows: comparison of the temperature, LOS velocity and magnetic field strength between the actual atmosphere (solid lines) and the inferred stratifications (dashed lines). The inferred atmospheres are plotted for the inversion with lowest $\chi^2$ (thick dashed lines), and the uncertainties are indicated by the thin dashed lines. Vertical dotted lines serve as references for some of the optical depths discussed in the text.}    
  \label{fig:inversion_tfijo}
\end{figure*}

\subsubsection{Inversions}
\label{sect:inversion_fix}

The spectral sampling of the synthesis illustrated in \hbox{Fig. \ref{fig:sintesis_tfijo}} is 5 m\AA. Since our aim is to evaluate the spectropolarimetric profiles in observing-like conditions, we have assumed a Gaussian spectral filter transmission. The profiles have been convolved with a Gaussian with FWHM of 100 m\AA. Since the main features of the unfiltered profiles are still visible in the degraded cases, we estimate that the influence of the spectral filter of the results of our analysis is minor. In the forthcoming analyses, only some selected wavelengths are retained in the profiles, in order to mimic the spectral positions acquired in standard observations taken by 2D instruments. We have chosen to reproduce the wavelength coverage from \citet{delaCruz-Rodriguez+etal2013}. The \CaII\ 8542 \AA\ line was sampled between $\Delta\lambda=-300$ m\AA\ and $\Delta\lambda=300$ m\AA\ in steps of 75 m\AA, with $\Delta\lambda$ defined as the wavelength difference with the core of the line, in addition to points at $\Delta\lambda=\pm 400,\pm 860,+3100$ m\AA. This set up provides an accurate sampling of the core of the line (with longer integration time) and a coarser sampling at the wings, allowing an efficient and fast acquisition of the full line profile. Using CRisp Imaging SpectroPolarimeter \citep[CRISP][]{Scharmer+etal2008}, a complete spectropolarimetric scan takes approximately 16 s \citep{delaCruz-Rodriguez+etal2013}.

The four Stokes profiles with the observational wavelength coverage were inverted with NICOLE for all the stages of an UF. The top panels of Figure \ref{fig:inversion_tfijo} show the synthetic Stokes $I$ and $V$ profiles during the development of the UF (at the same time steps illustrated in Fig. \ref{fig:sintesis_tfijo}) after applying the Gaussian spectral filter transmission at the chosen wavelengths. The thick dashed lines in the bottom panels of Figure \ref{fig:inversion_tfijo} show the atmospheric stratification retrieved from the best inversion of the instantaneous Stokes profiles, while the thin dashed lines mark the spread in the solutions defined by the standard deviation of five inversions with low $\chi^2$. The inferred atmospheric stratification is able to produce an accurate characterization of the spectral features showed by the instantaneous profiles, including the reversal of the intensity core, the associated polarity change in Stokes $V$, and the complex profiles that take place during this transition. The inferred atmospheric stratification shows a good agreement with the actual values (extracted directly from the simulation). Their comparison reveals the following insights:

\emph{Temperature}: The temperature of the model shows an approximately constant value around 4000 K with a sudden increase at the transition region. The location of the transition region is shifted progressively to lower heights as the shock wave generating the UF reaches the higher layers, from $\log\tau=-5.2$ at $t_1=507$ to $\log\tau=-4.7$ at $t_4=528$ (see vertical dotted lines). This displacement is captured by the inversions. Two different kind of solutions are found for the temperature. Some of them show a fast increase with height preceded by a dip with lower temperature than that from the actual model, while other solutions exhibit a temperature stratification in quantitative agreement with the simulation between the deep photospheric layers and approximately $\log\tau=-4.0$ followed by a smooth temperature enhancement. This mismatch is not a failure of the inversion process. A perturbation in the atmosphere in a region shorter than the photon mean free path would produce no changes to the radiation. This physical property of the radiative transfer limits the capability of inversion to determine variations over a thin layer.

\emph{LOS velocity}: Similarly to the inferred temperature, the velocity obtained from the inversion does not capture the sudden changes of the simulation. They correctly reproduce the sign of the velocity from the model at around $\log\tau=-5.0$, and provide a reasonable estimation of the amplitude. At $t_4=528$, the inverted velocity stratification shows a remarkable quantitative agreement in the amplitude of the upflow velocity, but the height of the velocity maximum is shifted to deeper layers. Interestingly, at $\log\tau=-5.5$ the inversion apparently reproduces the positive velocity of the model, although at that height the line is barely sensitive to the velocity (Fig. \ref{fig:RF}). Since the inferred atmosphere is characterized by a set of equidistant nodes, it is possible that the velocity results at the high atmosphere (above $\log\tau=-5.5$) are caused by a extrapolation of the lower heights. In order to verify this issue, we have performed an independent experiment. The synthesis of the simulation at that time step has been calculated again, but eliminating the atmosphere above $\log\tau=-5.2$. Thus, the atmosphere generating the synthetic Stokes profiles in this case has no positive velocity at higher layers. The inversion of these new profiles was computed following the same procedures described above, and the inferred stratification is very similar to that obtained previously. This way, we confirmed that the positive velocity inferred at the high layers is an artifact of the inversion due to the extrapolation from the slightly lower layers.

\emph{Magnetic field}: The inversion provides reasonably good estimates of the chromospheric vertical magnetic field from the simulation, except for the case at $t_2=514$ s which departs several hundred Gauss.

\begin{figure}[!ht] 
 \centering
 \includegraphics[width=9cm]{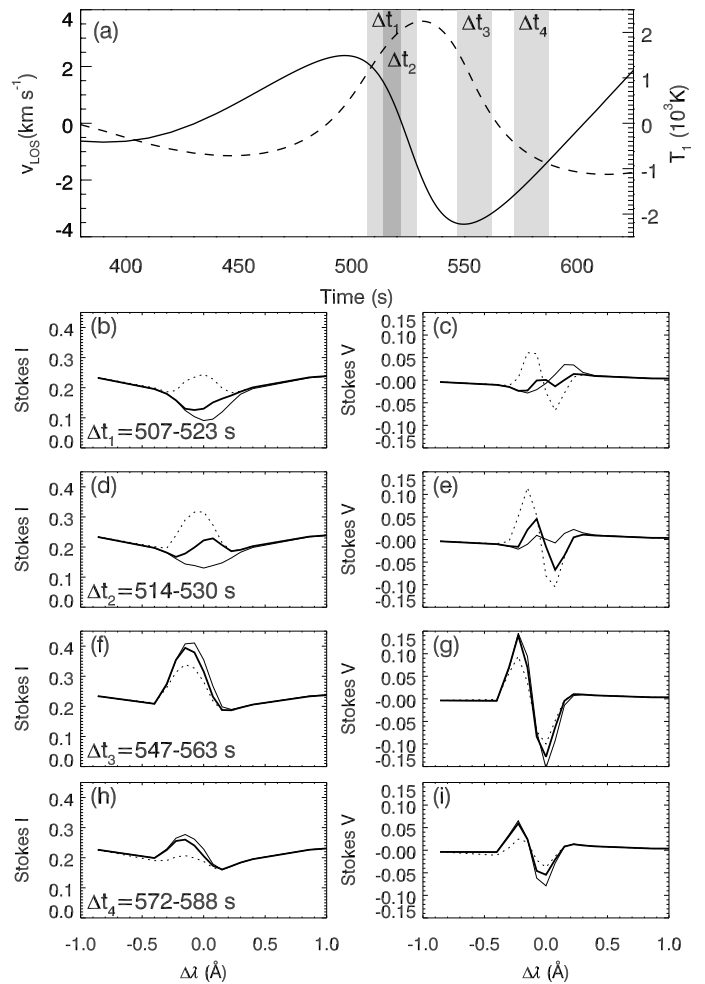}
  \caption{Spectropolarimetric synthesis of the \CaII\ 8542 \AA\ line imposing the time dependent acquisition of wavelengths in 2D instruments at several time ranges. Top panel: temporal evolution of the vertical velocity (solid line, left axis) and temperature perturbation (dashed line, right axis) at constant geometrical height \hbox{$z=750$ km}. Rest of the panels: Stokes I (left panels) and Stokes V (right panels) synthesized at the first time step from the range indicated at the top left part of the Stokes $I$ panel (thin solid line), at the last step of the range (thin dotted line), and as measured by 2D instruments (thick solid line). The time range of the acquisition of each profile is marked by the grey shaded areas in the top panel.}
    
  \label{fig:sintesis_tstep}
\end{figure}

\begin{figure*}[!ht] 
 \centering
 \includegraphics[width=18cm]{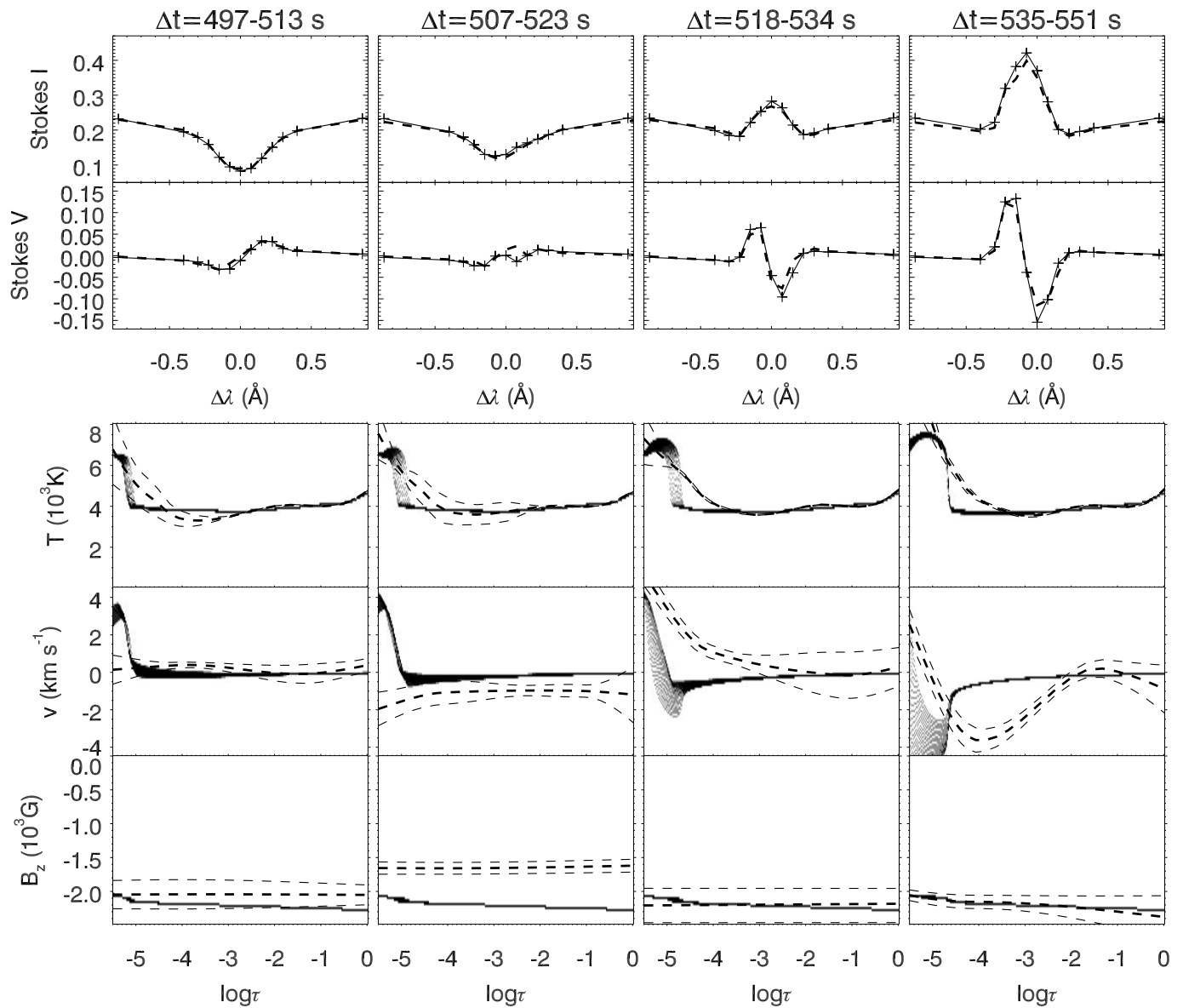}
  \caption{Same as Fig. \ref{fig:inversion_tfijo} but for synthetic Stokes profiles with wavelengths acquired at different time steps as obtained from 2D instruments (scanned profiles). The scanning imitates that of \citet{delaCruz-Rodriguez+etal2013}. The three bottom rows show density maps of all the atmospheres taking place at that position during the temporal span of the scanning (indicated at the top of the columns).}    
  \label{fig:inversion_tstep}
\end{figure*}

\subsection{Synthetic scanned observations}
\label{sect:tstep}

We have constructed synthetic Stokes profiles where each of the wavelengths is obtained at a different time step, in the same fashion that 2D instruments operate. We will refers to these profiles as ``scanned profiles''. We have chosen to emulate the observational set up from \citet{delaCruz-Rodriguez+etal2013}. The wavelength coverage is the same described in Sect. \ref{sect:fix_time}. The output of the numerical simulation was saved every 0.7 s. In this first approach, we assumed that the line profile is scanned from the blue side to the red side of the line. The Stokes profiles of the most blue wavelength ($\Delta\lambda=-400$ m\AA) were populated with the synthesis obtained from a single time step and for the following wavelengths we successively assigned the synthesis of the next time steps. At the wings (for spectral positions $\Delta\lambda=\pm 400, \pm 800, +3100$ m\AA) the time interval between consecutive line positions was 0.7 s, while at the core of the line (between $\Delta\lambda=\pm 300$ m\AA) is was 1.4 s. These intervals account for the effective integration time \citep[in the observations from][it was 210 ms at the wings and 900 ms at the core]{delaCruz-Rodriguez+etal2013} and the time required by the instrument for mechanical reasons (that we estimate as 0.5 s). We neglected the variation of the atmosphere during the integration time of each wavelength. The complete spectropolarimetric scan is obtained in 16.1 s, which is in agreement with the observational configuration from \citet{delaCruz-Rodriguez+etal2013}.

\subsubsection{Stokes profiles}

Figure \ref{fig:sintesis_tstep} shows the Stokes $I$ and $V$ profiles obtained following this approach at four time intervals during the development of the same UF described above. The first case ($\Delta t_1$) clearly illustrates the striking effect of the time-dependent acquisition of different wavelengths on the resultant profiles. At the beginning of the first scan ($t=507$ s, thin line from panel b), the core of the line is in absorption and shifted towards the red, according to the positive velocity (downflow, see top panel of Fig. \ref{fig:sintesis_tstep}) of the wave front. When the shock reaches the chromosphere, the core emission starts to arise. As the scanning of the line profile moves towards the red, the intensity increases (due to the development of the core reversal). As a result, the intensity profile still appears in absorption, but it is shifted towards the blue. That is, an atmosphere with a downward velocity produces an apparently blue shifted profile. This effect is an artifact of the scanning strategy. During the time interval $\Delta t_1$, the scanned Stokes $V$ shows several low amplitude lobes, similar to the instantaneous Stokes $V$ signal synthesized at the middle of the time interval $\Delta t_1$ (around $t=514$ s, see the thin line from panel e). However, the instantaneous intensity profile at $t=514$ s (thin line from panel d) shows clear differences with that of the scanned Stokes $I$ profile acquired during $\Delta t_1$. Even though scanned Stokes $V$ can be similar to some instantaneous profiles, they can potentially be misinterpreted since their corresponding Stokes $I$ are remarkably different.

The scanned Stokes $I$ profile obtained during $\Delta t_2$ (panel d) shows a developing UF. Interestingly, due to the same effect discussed in the previous paragraph, the reversed core is significantly redshifted. The instantaneous syntheses indicate that the center of the reversed core is slightly redshifted during a short time at the early stages of the flash. During the rest of the phenomena, it is clearly shifted towards the blue. Due to the progressive increase of the emission with time (and the scanning approach from blue to red), the red part of the core will exhibit higher intensity during the development of the UF, leading to a spurious redshift in the location of the core peak. The opposite situation takes place at the stage when the core emission is decreasing (scanned profiles in the time intervals $\Delta t_3$ and $\Delta t_4$) and the reversal is blueshifted. However, since at that phase the atmospheric changes are more gradual (in contrast to the simulation time between \hbox{$t=500$ s} and \hbox{$t=550$ s}, when the shock front reaches the formation height of the \CaII\ 8542 \AA\ line, Fig. \ref{fig:sintesis_tstep}a), the effects of the time-dependent scanning on the profiles are moderate. For simulation times prior to the arrival of the wavefront to the formation height of the \CaII\ 8542 \AA\ line (before \hbox{$t\sim 510$ s}) the differences between the scanned and instantaneous Stokes profiles are negligible.

\subsubsection{Inversions}

The inversion of the scanned Stokes profiles has been performed following the same strategy described in Sect. \ref{sect:methods}. Figure \ref{fig:inversion_tstep} illustrates the results of those inversions, including a comparison between the scanned Stokes $I$ and $V$ profiles and those obtained from the inversion with lowest $\chi^2$ (top two rows) and a comparison between the model atmospheric stratification (temperature, LOS velocity, and vertical magnetic field) and those inferred from the inversions (bottom three rows). The stratification of the simulation is represented by density plots of all the atmospheric models taking place during the time interval used for the construction of the scanned Stokes profiles (over \hbox{16.1 s} of simulation). In this temporal period the magnetic field hardly changes, but the temperature and LOS velocity show striking variations. The quality of the inversions depends on the phase of the UF:

\emph{Prior to the UF:} The inversion of the scanned Stokes profiles obtained in the time interval $\Delta t=497-513$ s (first column) provides a good inference of the actual atmosphere from the simulation. The temperature stratification captures the increase at the transition region and shows a small dip at lower heights, whereas the inferred LOS velocity is slightly positive, in agreement with most of the atmospheres at the maximum of the response function to velocity ($\log\tau=-5.0$). Note that during this time interval the effects of the time-depending scanning are barely visible.

\emph{During the development of the UF:} The scanned Stokes profiles constructed during $\Delta t=507-523$ s (second column) exhibit a spuriously blueshifted intensity profile. From an accurate fit of the intensity, the inversion code retrieves a relatively strong upflow ($\sim -1.5$ km s$^{-1}$ at $\log\tau=-5.0$) which is not present in any of the atmospheric models that generate the line profiles. The quality of the fit of Stokes $V$ is lower. Since this set of Stokes profiles is constructed from merging those produced by different atmospheres, the inversion code cannot find a single atmosphere that reproduces all the features of the profiles. The inferred vertical magnetic field clearly departs from the actual atmosphere. At the beginning of the flashed phase ($\Delta t=518-534$ s, third column), the peak of the emission core is located at $\Delta\lambda=0$ due to the spurious redshift introduced by the scanning process. Although the corresponding atmospheric models exhibit upflows with amplitudes as high as \hbox{-2 km s$^{-1}$}, the inversion code retrieves a strong downflow. This result is not due to the positive velocity that the actual atmospheric models show at higher layers (above $\log\tau=-5.2$), since the \CaII\ 8542 line is not sensitive to those heights. It is an artifact produced by the deformation of the Stokes profiles due to the sequential acquisition of different wavelengths.

\emph{Fully developed UF:} The fourth column from Fig. \ref{fig:inversion_tstep} shows a good fit of the Stokes profiles. In this phase, the time-dependent acquisition of the profiles only produces quantitative changes in the inferred atmosphere, but the stratification qualitatively reproduces the actual models. The previosly discussed spurious redshift of the emission peak during this stage may explain the lower velocity amplitude obtained from the inversion as compared with that from the simulation.

\begin{figure*}[!ht] 
 \centering
 \includegraphics[width=18cm]{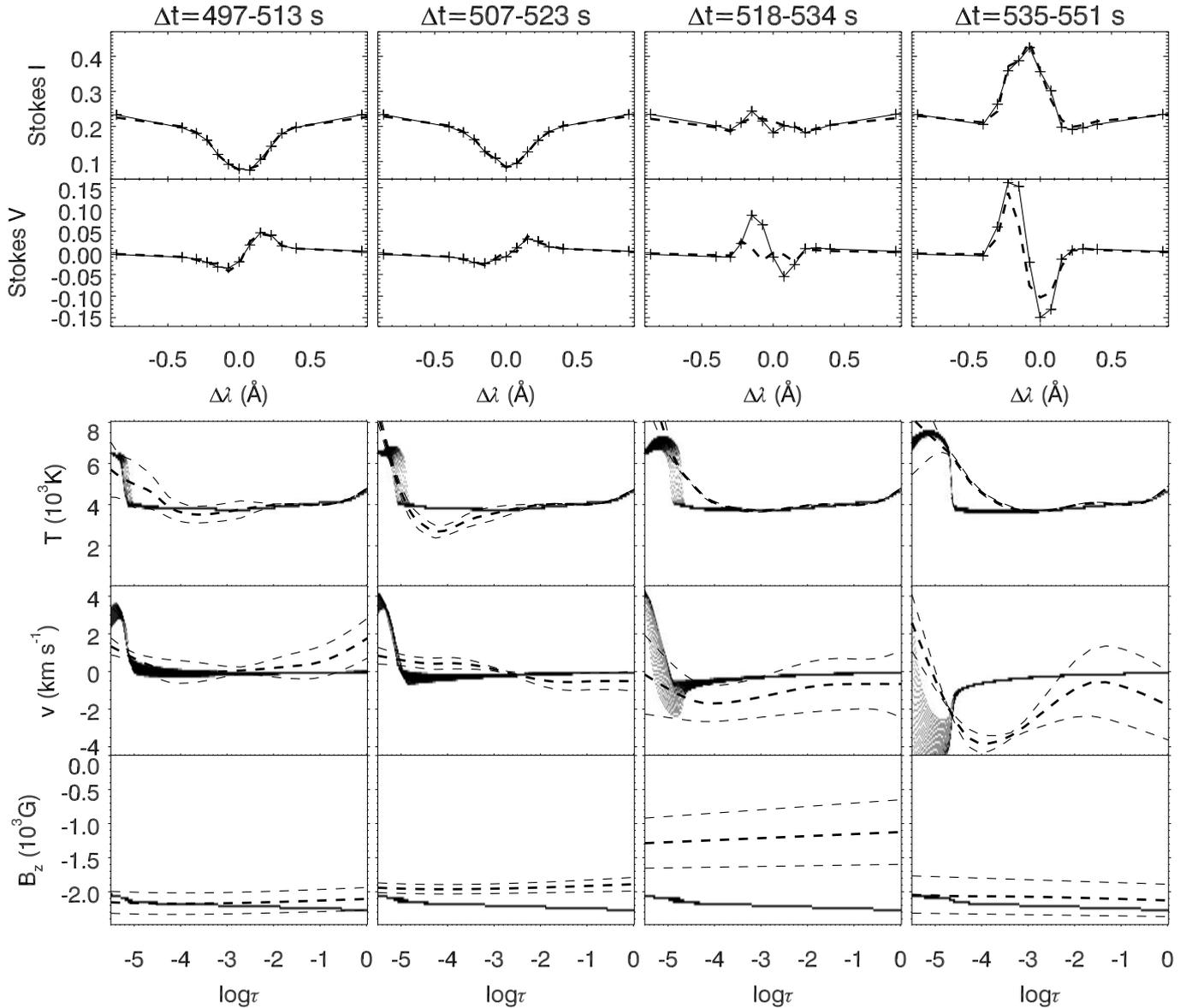}
  \caption{Same as Fig. \ref{fig:inversion_tstep}, but for a different acquisition strategy. The data covers the same wavelength positions from Figs. \ref{fig:inversion_tfijo} and \ref{fig:inversion_tstep} but the scanning starts from the core of the line and measures wavelengths at the red and blue side of the line successively. The temporal span of each scanning is indicated at the top of the columns. }    
  \label{fig:inversion_tstep_core}
\end{figure*}

The analysis of the inversion results has been repeated for a new set of scanned Stokes profiles. They were constructed using a different acquisition strategy. We selected the same wavelength coverage, but the order of the scanning was changed. Instead of sampling the line profile from the blue to the red, in this case we started from the center of the line $\Delta\lambda =0$ and then the wavelengths at the red and blue side were alternatively included. That is, the acquisition order is $\Delta\lambda =0$, +75, -75, +150, -150, +225, -225, +300, -300, +400, -400, +860, -860, +3100 m\AA. The same scanning time was assumed, with an interval of \hbox{1.4 s} at the core and \hbox{0.7 s} at the wings. This approach can potentially provide two advantages: (1) the signals from wavelengths with similar response functions are obtained close in time, and (2) the artificial wavelength shifts due to the sequential scanning of different wavelengths are minimized. One disadvantage is that the jumps in wavelength when scanning the wings are larger, and the etalons of Fabry-P\'erot interferometers require more time to stabilize. Generally, this method will need longer time cadences. The scanned profiles constructed following this scheme and the results of their inversions are illustrated in Fig. \ref{fig:inversion_tstep_core}.

The inversion of the first and last time intervals shown in Fig. \ref{fig:inversion_tstep_core} provides similar results to those obtained from the previous acquisition strategy (Fig. \ref{fig:inversion_tstep}). In contrast, the profiles retrieved from the early phases of the UF lead to significant differences in the inferred atmospheres. The inversion of the scanned profiles in the interval \hbox{$\Delta t=507-523$} s shows a small downflow (around 0.5 km s$^{-1}$ at $\log\tau=-5$). Two causes explain the difference between this result and that obtained from the other acquisition strategy (second column from \hbox{Fig. \ref{fig:inversion_tstep}}). First, although the time interval used for the construction of the profile is the same, in this case the wavelengths of the core were gathered at the beginning of the interval, when the upflow of the shock wave had not arrived to the formation height of the line. The velocity at the chromosphere corresponds approximately to \hbox{$t=507$ s}. This situation differs from the case shown in the second column from \hbox{Fig. \ref{fig:inversion_tstep}}, when the core was scanned around \hbox{$t=515$ s}. At that time, some atmospheric layers already exhibit negative velocity, as shown in the density plot of the simulated models. Second, the scanning method of \hbox{Fig. \ref{fig:inversion_tstep_core}} does not produce the spurious blueshift in the position of the line minimum and, thus, the inversion does not yield an unrealistic upflow.

When the core reversal is taking place, the scanned profiles (obtained by first scanning the core) exhibit a complex shape (third column from Fig. \ref{fig:inversion_tstep_core}). At this phase, the core emission increases with time. Since the wavelength position $\Delta\lambda =0$ is first measured, the surrounding wavelengths show a higher intensity than the central position, and the emission peak consist of two maximums separated by a minimum at $\Delta\lambda =0$ (top panel of third column from Fig. \ref{fig:inversion_tstep_core}). The inversion code is not able to reproduce this feature, and it also generates a poor fitting of the Stokes $V$. This is not a failure of the inversion method. There is simply no single atmospheric stratification that can account for the spurious features of the profiles. The inferred temperature and LOS velocity show a reasonable agreement with the actual models from the simulation, whereas the vertical magnetic field strongly departs from them. However, due to the low quality of the fits, any result obtained from the inversion of these profiles must be interpreted with care.

\subsection{Comparison of the inferred atmospheres}
\label{sect:comparison}

\begin{figure}[!ht] 
 \centering
 \includegraphics[width=9cm]{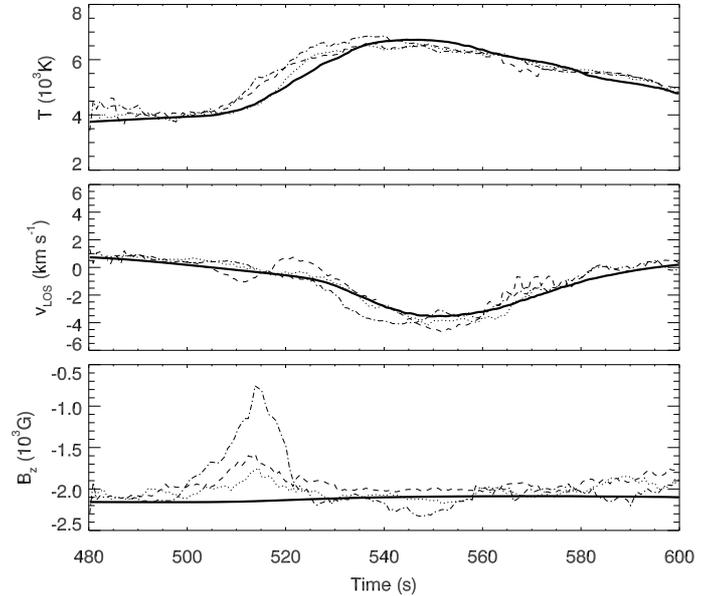}
  \caption{Temporal evolution of the temperature (top panel), LOS velocity (middle panel), and vertical magnetic field (bottom panel) at the chromosphere. Each panel illustrates the variable from the simulation (thick solid line), and the values inferred from the inversion of instantaneous profiles (dotted line), scanned profiles acquired from blue to red (dashed line), and scanned profiles acquired starting at the core (dashed-dotted line).}    
  \label{fig:evolution_chromosphere}
\end{figure}

Figure \ref{fig:evolution_chromosphere} illustrates a comparison between the chromospheric values of temperature, velocity, and magnetic field inferred from the inversions and those from the simulation. As previously discussed, the inversion results are not sensible to a discrete height. In order to perform a meaningful comparison, the values shown for the temperature and vertical magnetic field represent an average over optical depths between $\log\tau=-4.6$ and $\log\tau=-5.1$. The velocity from the simulation was averaged between $\log\tau=-4.6$ and $\log\tau=-4.9$, while the inverted velocities correspond to the mean between $\log\tau=-4.0$ and $\log\tau=-4.4$. For the velocity results we chose a lower height since, as previously shown, our inversions recover a good estimation of the velocity amplitude but shifted towards lower layers, probably due to the number and location of velocity nodes. Each point is obtained as the average of five time steps to minimize spurious fluctuations. For the inversions of scanned profiles, the horizontal axis indicates the scanning time of the wavelength position $\Delta\lambda =0$.

The temperature inferred from all the inversions shows a good agreement with the actual time variation from the simulation. The inversions capture the sudden temperature enhancement followed by a gentle decrease. The main differences between the three inversions are found during the time range when the temperature rises. The timing of the temperature increase retrieved from the instantaneous profiles (dotted line) is more faithful to that exhibited by the simulation (thick solid line) than those obtained from the scanned profiles (dashed and dashed-dotted lines).

In the case of the velocity, as the UF develops the chromospheric velocity changes progressively from a downflow to a strong upflow. Between $t=505$ s and $t=530$ s, the profiles scanned from blue to red (dashed line) significantly depart from the original value due to the wavelength shifts discussed in Sect. \ref{sect:tstep}. As the depth of the line core decreases the position of the intensity minimum is blueshifted and, thus, the inferred velocity is more negative (dashed line at $t=510$ s), whereas when the core is reversed its peak is redshifted and the estimated velocity is more positive (dashed line at $t=520$ s). Some departures between the inferred velocity and the simulation are also found at the maximum velocity of the upflow. 

The UFs are produced by slow waves propagating along field lines in a region dominated by magnetic field. The behavior of these waves is similar to acoustic waves and, at fixed geometrical height, they barely produce changes in the magnetic field. During the analyzed UF event, the vertical magnetic field variation at $z=750$ km is below 1.5 G. However, due to the strong changes in the opacity associated to the shock propagation and the consequent variation in the formation height of the line, the magnetic field at constant optical depth exhibits fluctuations slightly below 100 G (thick solid line in the bottom panel of Fig. \ref{fig:evolution_chromosphere}). None of the inverted atmospheres capture the changes in the magnetic field. The inversion of the instantaneous profiles shows an unphysical $\sim$300 G reduction of the field strength during the development of the UF. The inferred field is minimum at around $t=515$ s. This spurious variation in the magnetic field is enhanced in the scanned profiles. For the profiles which were acquired starting at the core, the inversion returns a magnetic field decrease of almost 1400 G. As shown in the third column of Fig. \ref{fig:inversion_tstep_core}, during this time the fitting of the profiles is fairly poor, and the inferred atmospheric model is unreliable.

Between $t=530$ s and $t=580$ s, the inversion of the instantaneous profiles (dotted line) shows a reasonable agreement with the values from the simulation. With regards to the scanned profiles, some mismatches are evident during this period.

\section{Discussion and Conclusions}
\label{sect:conclusions}

In this paper, we present the analysis of artificial \CaII\ 8542 \AA\ spectropolarimetric data generated from numerical simulations of non-linear wave propagation in a sunspot umbra. The synthesis naturally produces UFs, which are characterized by emission in the line core and polarity changes in Stokes $V$. After degrading the spectral resolution and reducing the wavelength coverage to that employed in actual observations \citep{delaCruz-Rodriguez+etal2013}, two kinds of synthetic spectropolarimetric data have been inverted with the NLTE code NICOLE: (1) instantaneous profiles, with all the wavelengths obtained at a single time step; and (2) scanned profiles, which are constructed with time-dependent acquisition of different wavelengths, simulating the scanning of 2D instruments.

The analysis of the instantaneous profiles prove that the inversion code can reasonably infer the atmospheric stratification of the simulated UF (Fig. \ref{fig:inversion_tfijo}), except for the magnetic field at the early stages of the UF. The velocity and temperature are recovered showing a good agreement with the chromospheric values of the simulation, although the inversion code approximates the abrupt changes associated to the shock with steep gradients. The main limitation is that the radiative transfer is sensitive to a range of physical heights, resulting in a ``blurring out'' of the structure in the vertical direction. Some information could be retrieved if one has a sufficiently high spectral resolution and/or many spectral lines with different but overlapping formation heights in the region of interest. However, one has an effective ``vertical resolution'' given by the information content of the observations. In our work, the inferred temperature cannot reproduce the sharp increase that the simulated model exhibits at the transition region. Instead, two different solutions are found. In some cases, the temperature enhancement is smoothed, and the inferred atmospheres show a progressive temperature increase starting at around $\log\tau=-4$. In other cases, the profiles are fit with a sudden temperature increase located above a dip with a reduced temperature value.   

The inversion of the scanned profiles shows similarly good results for the data acquired after the UF is fully developed and also for the profiles obtained prior to the UF. However, during the early stages of UFs, a visual inspection of the scanned profiles reveals features produced by the short characteristic time of the changes in the Stokes profiles as compared with the temporal cadence of the data acquisition. The main discrepancies are found during the period when the temporal derivatives of the temperature and velocity are strong. In the simulated case, this stage encompasses a time range of approximately 30 s just before the flashed emission peak reaches its maximum value. Since UFs are related to wave phenomena with a period around 180 s, this result indicates that around 15\% of the profiles in flashing regions may be affected by this issue.  
During this time, scanned Stokes $V$ can exhibit a profile similar to those synthesized at a single time step, but associated with a remarkably different intensity profile. This mismatch complicates the interpretation of spectropolarimetric spectra acquired by 2D instruments during dynamic events.

At the first stages of an UF, the intensity of the core of the line increases until producing a full reversal followed by the appearance of a prominent intensity peak. Since the Stokes signals at each wavelength are gathered at different times, the resulting scanned line profiles do not reproduce those generated by a single atmospheric model. When the scanning is performed from blue to red wavelengths (Fig. \ref{fig:sintesis_tstep}), the scanned Stokes $I$ profiles exhibit some deformations, such as a blueshift in the intensity minimum prior to the UF or a redshift in the emission peak during the development of the flash. These features are captured by the inversion code, which recovers an atmospheric stratification that departs from the original atmosphere. One of the most prominent manifestations of this fact is illustrated in the third column from Fig. \ref{fig:inversion_tstep}. In this case, the code recovers a strong downflow from the inversion of an UF, even though the actual velocity of the model mainly exhibits an upflow for the optical depths where the \CaII\ 8542 \AA\ line is sensitive to the velocity. Interestingly, a strongly downflowing chromosphere in UFs has recently been claimed for the first time \citep{Henriques+etal2017}. The authors propose that these chromospheric downflows (if real) may be related to the presence of coronal loops and the associated downflows above the umbra reported by many previous works \citep[\eg,][]{Dere1982, Vissers+RouppevanderVoort2012,Kleint+etal2014,Chitta+etal2016}. Our results show an alternative explanation for the detection of downflowing UFs. They can be retrieved as an artifact of the data gathering process. However, some differences between our modeling and \citet{Henriques+etal2017} observations must be noticed. Firstly, the emission peak of the spurious downflowing UFs obtained from the scanned Stokes $I$ profiles (scanning from blue to red) is located at the center of the line or slightly redshifted, whereas \citet{Henriques+etal2017} obtained chromospheric downflows from the inversion of some flash profiles with blueshifted emission core. Secondly, the time dependence of the wavelength acquisition is not directly comparable. Each analyzed scan from \citet{Henriques+etal2017} data is obtained from the reconstruction of two observed scans using MOMFBD. The total acquisition time is 28 s, significantly longer than the 16.1 s cadence used in this work. Our analysis cannot discard the real nature of the downflowing UFs, but points out that some artifacts produced by the data gathering process of 2D instruments can lead to a misinterpretation of the observations.      

Finally, it is also noted that the magnetic field recovered from the inversion of the Stokes profiles can significantly depart from the actual atmospheric values during the development of UFs (when the shock wave reaches the formation height of the \CaII\ 8542 \AA), producing spurious magnetic field variations of several hundred Gauss in the inferred field strength. This issue is even more critical for the scanned profiles.

\begin{acknowledgements} 
The authors thank Jaime de la Cruz Rodr\'iguez for his helpful comments to an early version of the manuscript. Financial support from the Spanish Ministry of Economy and Competitivity through projects AYA2014-55078-P, AYA2014-60476-P and AYA2014-60833-P is gratefully acknowledged. The authors wish to acknowledge the contribution of Teide High-Performance Computing facilities to the results of this research. TeideHPC facilities are provided by the Instituto Tecnol\'ogico y de \hbox{Energ\'ias} Renovables (ITER, SA). URL: http://teidehpc.iter.es. This work was also supported by computational resources provided by the Australian Government through the Pawsey Supercomputing Centre under the National Computational Merit Allocation Scheme.
\end{acknowledgements}

\bibliographystyle{aa} 
\bibliography{biblio.bib}

\begin{appendix}

\section{MHD simulations}
\label{appendix:simulations}

Numerical simulations of wave propagation from the photosphere to higher atmospheric layers in a sunspot were performed using the code MANCHA \citep{Khomenko+Collados2006, Felipe+etal2010a}. The code solves the non-linear equations for perturbations, which are obtained after explicitly removing the equilibrium state from the equations. Periodic boundary conditions were imposed in the horizontal directions and perfect matched layers \citep{Berenger1996} were used in the top and bottom boundaries in order to damp waves with minimum reflection.

Waves are driven by adding spatially-localized sources of vertical force in the equations. Their temporal variation is described in \citet{Parchevsky+etal2008}, and generates a photospheric power spectra that reproduces that of the Sun \citep{Parchevsky+etal2008, Felipe+etal2016a}. The sources were located 0.2 Mm below the photosphere. They generate a (non-linear) perturbation to a magnetohydrostatic (MHS) background model. The sunspot model was constructed according to \citet{Przybylski+etal2015}. It is based on the method by \citet{Khomenko+Collados2008}, but it was modified in order to provide an accurate characterization of line formation regions, allowing detailed spectropolarimetric synthesis. Semi-empirical models are required at the axis of the sunspot and at the quiet Sun region as boundary conditions for the construction of the MHS sunspot. At the quiet Sun, a convectively stable \citep[following][]{Parchevsky+Kosovichev2007} Standard Solar Model S \citep{Christensen-Dalsgaard+etal1996} was imposed in the deep layers, and it was smoothly joined with VAL-C model \citep{Vernazza+etal1981}. For the inner boundary atmosphere (axis of the sunspot), a modified version of \citet{Avrett1981} umbral model was employed. The synthesis of the \CaII\ 8542 \AA\ line in the Avrett model shows a line core emission reversal, which is not observed in a quiescent umbral atmosphere \citep[\eg,][]{delaCruz-Rodriguez+etal2013}. In order to obtain an accurate characterization of the expected line profile for the umbra at rest, the transition region has been shifted 200 km higher when compared to the original Avrett model. The constructed MHS sunspot has a Wilson depression of 450 km and an umbral photospheric magnetic field strength of 2300 G. The adiabatic index is obtained from the OPAL equation of state \citep{Rogers+etal1996}.

In this work, for simplicity we use a 2.5D approximation. All the vectors are considered in the three spatial directions, but the derivatives are only calculated in the vertical and one horizontal direction. This way, the perturbations are restricted to the $X-Z$ plane. The computational domain has an horizontal extent of 140 Mm with a resolution of \hbox{$\Delta x=0.2$ Mm}. The axis of the sunspot is located at the center position. The vertical direction spans from \hbox{$z=-4.65$ Mm} to \hbox{$z=2$ Mm}, with the height $z=0$ defined at the layer where the optical depth at \hbox{5000 \AA\ ($\tau$)} is unity in the quiet Sun. The code uses a grid with variable step size in the vertical direction based on the quiet-Sun sound speed. The acoustic travel time between adjacent vertical cells is the same, except for the higher layers above the temperature minimum (sound speed minimum), where the finest resolution (that of the temperature minimum) is forced. This way, a better sampling of the solar atmosphere is achieved, which allows a proper characterization of the shock waves. A maximum vertical step of \hbox{$\Delta z=44$ km} is used at the lower boundary, while in the photosphere and higher atmospheric regions \hbox{$\Delta z=15$ km}. The size of the numerical grid is $700\times 288$.

\end{appendix}

\end{document}